\pdfoutput=1
\documentclass{nice-article}

\usepackage[
  backend=biber,
  natbib=true,
  doi=true,
  citestyle=alphabetic,
  backref=true,
  sortcites=true,
  style=alphabetic,
  maxnames=1000,
  mincrossrefs=99,
  sorting=ynt,
  sortcites
]{biblatex}

\usepackage[scr]{rsfso}

\usepackage{dieudonne}

\usepackage{newtxmath}
\usepackage{newtxtext}

\RenewDocumentCommand\DieudonneInterpunct{}{\textasteriskcentered}
\usepackage{zi4}

\usepackage{macros}

\crefname{axiom}{Axiom}{Axioms}

\title{The directed plump ordering}
\author{Daniel Gratzer\and Michael Shulman\and Jonathan Sterling}

\hypersetup{
  pdfauthor = {Daniel Gratzer and Michael Shulman and Jonathan Sterling},
  pdftitle = {The directed plump ordering},
}

\begin{document}

\maketitle

\begin{abstract}
  Based on Taylor's hereditarily directed plump ordinals, we define the
  \emph{directed plump ordering} on W-types in Martin-L{\"o}f type theory. This ordering is similar
  to the plump ordering but comes equipped with non-empty finite joins
  in addition to the usual properties of the plump ordering.
\end{abstract}

\CounterZeroNext{node}

\begin{node}
  \emph{Acknowledgment.}
  This research was supported by the United States Air Force Office of
Scientific Research under award number FA9550-21-1-0009.
\end{node}

\begin{node}
  \label{node:intro}
  The theory of plump ordinals~\citep{taylor:1996} has been adapted to Martin-L{\"o}f type theory by
  \citet{fiore-pitts-steenkamp:2021} to produce directed well-founded orders suitable for certain
  transfinite constructions. Given a pair $\prn{A : \TY_1, B : A \to \TY_1}$,
  \emph{op. cit.} defines the \emph{plump ordering}: a pair of relations $\le,\Covered$ on a type $W$
  of well-founded trees satisfying the following conditions:
  \begin{enumerate}
  \item $\le$ is reflexive and transitive
  \item $\Covered$ is transitive and well-founded.
  \item If $u \Covered v$ then $u \le v$.
  \item If $u \Covered v \le w$ or $u \le v \Covered w$ then $u \Covered w$.
  \item $(W,\le)$ has a least element.
  \item For each $a : A$, both $\le$ and $\Covered$ have upper-bounds for all $B\prn{a}$-families.
  \end{enumerate}

  Following Taylor's theory of hereditarily directed plump ordinals~\citep{taylor:1996}, we refine
  this ordering to obtain well-behaved least upper-bounds:
  \begin{enumerate}[resume]
  \item Given $u,v : W$ there exists $u \Lub v$ such that $u \Lub v \le w$ if and only if
    $u,v \le w$.
    \label{item:lub}
  \item If $u, v \Covered w$ then $u \Lub v \Covered w$.
    \label{item:directed}
  \end{enumerate}
\end{node}

\begin{node}
  We have partially formalized our results in Martin-L\"of type theory with the
  UIP principle in the Agda proof
  assistant~\cite{agda-directed-plump-ordering}.\footnote{\url{http://www.jonmsterling.com/agda-directed-plump-ordering/}.}
  In particular, all results except the well-foundedness of the list ordering
  $\sqsubset$ of \cref{sec:list-ordering} are formalized in Agda.
\end{node}

\section{An ordering on W-types}

\begin{node}
  Fix a $\TY_1$-container $A\rhd B$ in the sense of
  \citet{abbott-altenkirch-ghani:2005}, \ie a pair of a type $A:\TY_1$ together
  with a family of types $B : A \to \TY_1$. The \emph{extension} of $A \rhd B$
  is the endofunctor $\bbrk{A\rhd B} : \TY_1\to\TY_1$ defined like so:
  \[
    \begin{make-rcd}{\bbrk{A\rhd B}}[X:\TY_1]{\TY_1}
      \mathbf{constructor}\ \prn{-,-}
      \\
      \TAG : A
      \\
      \CHLD : B(\TAG) \to X
    \end{make-rcd}
  \]

  The extension of a container is also known as the \emph{polynomial
  endofunctor} associated to the corresponding morphism
  $\Mor{\Sum{x:A}B\prn{x}}{A}$.
\end{node}

\begin{node}
  The \emph{initial algebra} for the extension $\bbrk{A\rhd B}$ of a given
  container can be computed as a W-type in the sense of \citet{martin-lof:1984}
  consisting of well-founded trees labeled in $a:A$ with subtrees of arity
  $B\prn{a}$, written $\W{A}{B}:\TY_1$. The structure map for this initial
  algebra is written $\Mor[\UB]{\bbrk{A\rhd B}\prn{\W{A}{B}}}{\W{A}{B}}$, which
  can be thought of as producing an upper-bound in the subtree order.
\end{node}

\begin{node}
  Suppose that the container $A\rhd B$ is closed under binary coproducts of shapes in
  the sense that we have an operation $\hatplus : A \times A \to A$ such that
  $B\prn{a_1\hatplus a_2} = B\prn{a_1} + B\prn{a_2}$.  Given two trees
  $u,v:\W{A}{B}$, we will write $u\Lub v$ for $\UB\prn{u.\TAG\hatplus v.\TAG,
  \Split{u.\CHLD}{v.\CHLD}}$. For a non-empty finite set of trees
  $\Compr{u_i}{i\leq n}$, we will write $\BigLub{i}{u_i}$ for the corresponding
  $n$-ary instance of $\Lub$.
\end{node}

\begin{node}
  We may define the following two
  binary relations $\le,\Covered$ on $\W{A}{B}$ as the
  smallest ones closed under the following rules:
  \begin{mathpar}
    \inferrule{
      \exists b_1, \dots b_n : B(v.\TAG).\
      u \le \BigLub{i}{v.\CHLD{b_i}}
    }{
      u \Covered v
    }
    \and
    \inferrule{
      \forall{b : B(u.\TAG)}.\ u.\CHLD(b) \Covered v
    }{
      u \le v
    }
  \end{mathpar}
\end{node}

Each of \cref{lem:le-refl,lem:trans-and-flex,lem:covered-to-le,lem:lub-universal-property} has been
formally verified in Agda.
\bigskip

\begin{node}\label{lem:le-refl}%
  The relation $\le$ is reflexive.
\end{node}

\begin{node}\label{lem:trans-and-flex}%
  For any $u,v,w:\W{A}{B}$ we have the following:
  \begin{enumerate}
    \item \emph{Transitivity.} If $u \le v \le w$ then $u \le w$; likewise if $u\Covered v \Covered w$ then $u\Covered w$.
    \item \emph{Left flex.} If $u \le v$ and $v \Covered w$ then $u \Covered w$.
    \item \emph{Right flex.} If $u \Covered v$ and $v \le w$ then $u \Covered w$.
  \end{enumerate}
\end{node}

\begin{node}\label{lem:covered-to-le}%
  For any $u,v:\W{A}{B}$, if $u \Covered v$ then $u \le v$.
\end{node}

\begin{node}\label{lem:lub-universal-property}%
  Let $\Compr{u_i}{i\leq n}$ be a non-empty finite family of trees, and let
  $v:\W{A}{B}$ be a tree; we have $\BigLub{i}u_i \le v$ if and only if $u_i\le
  v$ for all $i\leq n$. Morever, we have $\BigLub{i}{u_i}\Covered v$ if
  $u_i\Covered v$ for all $i\leq n$.
\end{node}

\section{An intermezzo on list orderings}\label{sec:list-ordering}

\begin{node}
  Given a relation $\Mor[R]{A \times A}{\Prop}$, define the accessibility predicate as the following
  inductive type:
  \[
    \begin{make-data}{\Acc{R}}{A \to \Prop}
      \Con{acc} : (a : A) \to ((b : A) \to R(b,a) \to \Acc{R}{b}) \to \Acc{R}{a}
    \end{make-data}
  \]
  A relation is said to be well-founded when all its elements are accessible. Note that a
  well-founded relation need not be transitive.
\end{node}

\begin{node}
  We eventually wish to show that $\Covered$ is well-founded but prior to this we must introduce a
  supplementary well-founded ordering. The well-foundedness of $\Covered$ will follow from
  well-founded induction on this secondary ordering.

  Fix a type $X$ and a well-founded relation ${<} : X \times X \to \Prop$ for the remainder of this
  section. We define a new relation $\sqsubset$ on $\List{X}$:
  \[
    \inferrule{
      m \ge 1
      \\
      \exists f : \brc{1 \dots n} \to \brc{1 \dots m}.\
      \forall i \le n.\ x_i < y_{f\prn{i}}
    }{
      \brk{x_1, \dots, x_n} \sqsubset \brk{y_1, \dots, y_m}
    }
  \]

  We adapt a proof due to Wilfried Buchholz as described by \citet{nipkow:1998}
  to prove that $\sqsubset$ is well-founded.
\end{node}

\begin{node}\label{lem:sqsubset-wf-base}%
  The empty list is $\sqsubset$-accessible.
\end{node}

\begin{node}
  If a list is $\sqsubset$-accessible, so too is any permutation.
\end{node}

\begin{node}\label{lem:sqsubset-wf-ind-helper}%
  Fix $y : X$. Suppose for all accessible $l : \List{X}$ and $x < y$, $\Cons{x}{l}$ is
  accessible. Then for all accessible $l : \List{X}$, $\Cons{y}{l}$ is accessible.
  \begin{proof}
    Fix an accessible $l$ and suppose that $n \sqsubset \Cons{y}{l}$. By definition, there exists a
    division of $n$ into $n_l$ and $n_y$ such that $n_l \sqsubset l$ and each element of $n_y$ is
    dominated by $y$. Because $l$ is accessible, so too is $n_l$. Therefore, $n_y + n_l$ is
    accessible by induction on the size of $n_y$ and repeated use of the assumption. Because $n$ is
    a permutation of $n_y + n_l$, we conclude that $n$ is accessible.
  \end{proof}
\end{node}

\begin{node}\label{lem:sqsubset-wf-ind}%
  If $l : \List{X}$ is $\sqsubset$-accessible and $x : X$, then $\Cons{x}{l}$ is accessible.
  \begin{proof}
    This follows immediately from the \cref{lem:sqsubset-wf-ind-helper} and $<$-induction on $x$.
  \end{proof}
\end{node}

\begin{node}
  If $<$ is well-founded, so too is $\sqsubset$.
  \begin{proof}
    Fix $l : \List{X}$. We argue by induction on $l$ that $l$ is accessible. In the
    base case apply \cref{lem:sqsubset-wf-base} and in the inductive step apply
    \cref{lem:sqsubset-wf-ind}.
  \end{proof}
\end{node}

\section{Well-foundedness of the directed plump ordering}

\begin{node}
  Write $\NEmptyList{X}$ for the type of \emph{non-empty} lists.
  Given an non-empty list $l = \brk{u_0, \dots, u_n}$, write $\BigLub{}{l}$ for
  $\BigLub{i\leq n}{u_i}$.
\end{node}

\begin{node}\label{lem:cov-wf-helper}%
  Given $l:\NEmptyList{\W{A}{B}}$, if $u\le \BigLub{}{l}$ then $u$ is
  $\Covered$-accessible.
  \begin{proof}
    This follows by well-founded induction on the $\sqsubset$-accessibility of
    $l$; the details are formalized in Agda.
  \end{proof}
\end{node}

\begin{node}
  The relation $\Covered$ is well-founded.
  \begin{proof}
    We must prove that every $u : \W{A}{B}$ is $\Covered$-accessible, but this
    is a consequence of \cref{lem:cov-wf-helper} setting $l$ to be the
    singleton list $\brk{u}$; the details are formalized in Agda.
  \end{proof}
\end{node}

\begin{node}\label{thm:directed-plump-ordinals}%
  Summarizing, given a pair $\prn{A : \TY_1, B : A \to \TY_1}$ together with an operation an
  operation $\hatplus : A \times A \to A$ such that
  $B\prn{a_1\hatplus a_2} = B\prn{a_1} + B\prn{a_2}$ there exists a type $\W{A}{B}$ together with a
  pair of relations ${\le},{\Covered} : \W{A}{B} \times \W{A}{B} \to \Prop$ satisfying the following
  conditions:
  \begin{enumerate}
  \item $\le$ is transitive and reflexive.
  \item $\Covered$ is transitive and well-founded.
  \item If $u \Covered v$, then $u \le v$.
  \item If $u \Covered v \le w$ or $u \le v \Covered w$ then $u \Covered w$
  \item If there exists $a : A$ such that $B\prn{a} = \InitObj{}$ then
    $\prn{\W{A}{B},{\le}}$ has a least element.
  \item For any $a : A$, both $\le$ and $\Covered$ have upper-bounds for all $B(a)$-families.
  \item Given $u,v$ there exists an element $u \Lub v$ such that $u \Lub v \le w$ if and only if
    $u,v \le w$.
  \item If $u,v \Covered w$ then $u \Lub v \Covered w$.
  \end{enumerate}
\end{node}

\begin{node}
  Given a pair $\prn{A : \TY_1, B : A \to \TY_1}$, define a new pair $\prn{C,D}$ by setting
  $C = \List{A}$ and specifying $D$ inductively:
  \[
    D\prn{ \brk{} } = \InitObj{}
    \qquad
    D\prn{ \Cons{a}{c} } = B\prn{a} + D\prn{c}
  \]

  Then \cref{thm:directed-plump-ordinals} instantiated with this new family shows that
  $\prn{\W{C}{D},\le,\Covered}$ satisfies the requirements outlined by \cref{node:intro}.
\end{node}

\printbibliography

\end{document}